# A Casimir approach to dark energy

**Allan Rosencwaig***

*Arist Instruments, Inc.*
*Fremont, CA 94538*

## Abstract

We calculate the gravitational self-energy of vacuum quantum field fluctuations using a Casimir approach. We find that the Casimir gravitational self-energy density can account for the measured dark energy density when the SUSY-breaking energy is approximately 5 TeV, in good agreement with current estimates. Furthermore, the Casimir gravitational self-energy appears to provide a quantum mechanism for the well-know geometric relation between the Planck, SUSY and cosmological constant energy scales.

* E-mail address: allan@aristinst.com

There is convincing evidence from supernovae and CMB studies that the decelerating universe in the past has evolved into an accelerating one at present [1-4]. This acceleration is apparently due to the presence of dark energy with negative pressure, that is, where the parameter $w$ in the equation of state that relates the pressure $p$ to the density $\rho$, $p = w\rho$, is < -1/3. Furthermore, this dark energy now appears to comprise about 72% of the total mass/energy density in the universe. There are two primary candidates for dark energy: a cosmological constant such as might arise from vacuum energy (quantum field vacuum fluctuations or zero-point energy) which results in a dark energy density that is constant with time and has $w = -1$ [5]; or quintessence as might arise from a time-varying scalar field which results in a dark energy density that decreases with time and has $-1 < w < -1/3$ [6,7]. Observations indicate that the present value of $w$ is close to -1 strongly suggesting that a cosmological constant is the best candidate.

The most natural explanation for dark energy is the vacuum or zero-point energy of space itself since it has a constant value and a $w = -1$. But the apparent magnitude of the vacuum energy is much too high. Vacuum fluctuations can be treated as the appearance and disappearance of virtual particles which exist for only a Planck time, with the most massive virtual particles contributing the most to the vacuum energy density. Thus, the greatest contribution to the vacuum energy density would come from the Planck particle with a mass of 2.2 x$10^{-5}$ g, giving a vacuum energy density of ≈ 5 x $10^{92}$ g/cm$^3$, or about 123 orders of magnitude greater than the presently accepted dark energy density of ≈ $10^{-29}$ g/cm$^3$. This is the well-known "cosmological constant problem".[5] It is possible to alleviate this difficulty somewhat by invoking supersymmetry [8]. At some early supersymmetric (SUSY) stage in the evolution of the universe, every fermion particle had a corresponding boson particle, and at some later stage, there occurred a spontaneous breaking of supersymmetry and the perfect balance between fermions and bosons was lost. If we invoke supersymmetry breaking we can define a SUSY-breaking mass $m_x$, ie a SUSY-breaking virtual particle, such that particles with mass smaller than $m_x$ would have no identical supersymmetric partners while those with greater mass would. Supersymmetry has a profound effect on the vacuum energy density because the vacuum energy from virtual fermions is negative while that from virtual bosons is positive.[5] Thus, if supersymmetry and supersymmetry breaking apply to virtual particles, then we only have to consider particles up to mass $m_x$ since the net vacuum energy density contributed by particles above mass $m_x$ is identically zero. Current estimates for the SUSY-breaking energy are in the TeV scale, which would indicate a vacuum energy density > $10^{33}$ g/cm$^3$. This reduces the cosmological problem somewhat but the SUSY-breaking vacuum energy is still more than $10^{62}$ times larger than the measured dark energy.

Although vacuum zero-point energy has not been detected directly, quantum electromagnetic field fluctuations do produce measurable effects, such as the Casimir energy [9], the Lamb shift [10] and the radiative correction to the electron magnetic moment [11]. The electromagnetic Casimir energy is the result of a local non-uniformity, such as a depression or cavity, in the electromagnetic configuration space. The non-uniformity in results in a different electromagnetic energy content inside the cavity as compared to outside. This difference between two infinite fluctuating energy fields is the Casimir energy, a finite non-fluctuating energy that is imparted to the cavity.
The electromagnetic Casimir energy is found to be attractive for most boundary configurations [12]. However, for the very important case of a perfectly conducting hollow sphere of radius $r$, it is found to be repulsive and is given by [13-15],

$$U_s(e/m) = 0.0462 \frac{\hbar c}{r} \qquad (1)$$

Similarly, when the vacuum quantum field is a fermionic Dirac field, the Casimir energy in a perfectly confining shell or cavity of radius $r$ is found to be [16],

$$U_s(Dirac) = 0.02 \frac{\hbar c}{r} \qquad (2)$$

In a previous paper we showed how a Casimir energy approach can be used to obtain good results for the one-photon radiative correction term to the Lamb shift for electrons in hydrogen orbitals [17]. We also showed how the Casimir energy approach can be used to calculate the electromagnetic energy of a free electron and the radiative correction to the electron gyromagnetic ratio [18]. The key element to our model is to extend the usual Casimir energy concept to a shell composed of only one electron and to include the probability of an interaction between the single electron and a virtual photon by means of the relativistic electromagnetic scattering cross-section. In the case where the electron is in a hydrogen orbital, we also include the probability of the electron being at a particular radial position in that orbital through the appropriate hydrogen wavefunction.

We found that the electromagnetic Casimir energy, $U_C(em)$, for a spherical shell of radius $r$ with one electron in the shell can be given by [18],

$$U_C(em) = \frac{1}{4\pi} \frac{\hbar c}{r} \frac{\sigma_T(em)}{r^2} = \frac{1}{2} \frac{\hbar c}{r} \frac{a_0^2}{r^2} \left\{ \frac{r}{\lambda_c} h\left(\frac{r}{\lambda_c}\right) \right\} \qquad (3)$$

and $$\sigma_T(em) = 2\pi a_0^2 \left\{ \frac{r}{\lambda_c} h\left(\frac{r}{\lambda_c}\right) \right\} = 2\pi a_0^2 h_1\left(\frac{r}{\lambda_c}\right) \qquad (4)$$

where $\sigma_T(em)$ is the total electromagnetic scattering cross-section, $a_0$ is the classical electron radius = $e^2/m_e c^2$ with $e$ the charge and $m_e$ the mass of the electron, $\lambda_c$ is the electron Compton wavelength = $2\pi \hbar /m_e c$, and $h$ is a function that incorporates the relativistic decrease of $\sigma_T$ with photon energy via a Klein-Nishina term.

For the case of an electron in a hydrogen orbital, the Casimir energy is the result of the non-uniformity in the electromagnetic configuration space of the orbital. For a free electron, which we assume is a quantum point particle, our picture would be of a local depression or cavity in the electromagnetic configuration space around the point charge of the electron. That is, the cavity or local depression in the local electromagnetic configuration space is caused by the charge of the electron and the Casimir energy arises from the interactions of the quantum vacuum electromagnetic field with this cavity.

We can also have the situation where the quantum field can produce its own local non-uniformity or cavity and then interact with it, that is, a self-energy term for the fluctuation itself. The easiest way to treat this problem is to consider the vacuum fluctuation as a momentary creation and annihilation of a virtual particle-antiparticle pair that survives only a Planck time. The momentary presence of such a particle pair will create a non-uniformity or cavity in the gravitational configuration space, that is, create a local spacetime curvature. A Casimir energy will then arise from the interactions of all of the other vacuum fluctuations with this local non-uniformity in the gravitational configuration.

Let us consider the non-uniformity as a shell of radius $r$. Then, in analogy with the case of an electron, a Casimir energy for this situation may be written as,

$$U_C(g) = C_g \frac{\hbar c}{r} \frac{\sigma_T(g)}{r^2} \qquad (5)$$
$$\sigma_T(g) = 2\pi a_g^2$$

We do not include in the gravitational scattering cross-section, $\sigma_T(g)$, an explicit term analogous to the Klein-Nishina term, $h_1(r/\lambda_c)$, in Eqn. (4) for two reasons. First, as we will see below, the gravitational cross-section inherently incorporates its own energy dependence. Secondly, for a quantum fluctuation $r/\lambda_c$ is always equal to $1/4\pi$, and thus $h_1(r/\lambda_c)$ is simply a number that is included in $C_g$.

In the electromagnetic case the scattering radius $a_0$ is given by,

$$a_0 = \alpha \lambdabar_c(e) \qquad (6)$$

where α is the fine structure constant or the electromagnetic coupling parameter and $\lambdabar_c(e)$ is the reduced Compton wavelength of the electron, $\lambdabar_c(e) = \lambda_c/2\pi$.

In analogy with Eqn. (6), we can write for the gravitational scattering radius, $a_g$, when the gravitational cavity is created by a vacuum fluctuation of mass $m$,

$$a_g = \alpha_g \lambdabar_c(m) = \frac{Gm^2}{\hbar c} \frac{\hbar}{mc} = \frac{Gm}{c^2} \qquad (7)$$

Thus the gravitational scattering radius is ½ the Schwartzschild radius. Note that whereas the electromagnetic scattering radius, $a_0$, is a constant and the energy dependence of the electromagnetic scattering cross-section is provided by the Klein-Nishina term, the gravitational scattering radius, $a_g$, is inherently dependent on energy and no additional energy dependent term is required. Also, whereas the electromagnetic scattering cross-section decreases with energy, the gravitational scattering cross-section increases with energy.

Now the radius $r$ in Eqn. (5) is simply $1/2\, \lambdabar_c(m)$. Thus the Casimir energy will be,

$$U_C(g) = 16\pi C_g \alpha_g^2 \frac{\hbar c}{\lambdabar_c} = 16\pi C_g \alpha_g^2 mc^2 \tag{8}$$

To obtain $C_g$ we consider the unique case where $m = m_{Pl}$, the Planck mass, at which point $a_g = 1$. At $m_{Pl}$ we expect $U_C(g) = m_{Pl}c^2$. Thus,

$$C_g = \frac{1}{16\pi} = 0.02 \tag{9}$$

We note that this is the same numerical factor that has been obtained for the shell Casimir energy for confined fermionic Dirac quantum fields (Eqn. (2)). We then obtain for any virtual pair of mass $m$,

$$U_C(g) = \alpha_g^2 mc^2 = \frac{G^2 m^3}{\hbar^2} \tag{10}$$

If there is no energy cut-off, then $U_C$ would simply increase up to $m_{Pl}c^2$, and we would have the same cosmological constant problem. However, in the energy region where supersymmetry is unbroken the Casimir energy, like the vacuum energy, is identically zero [19]. Thus we need evaluate $U_C$ only at the mass, $m_{SUSY}$, corresponding to the SUSY-breaking energy. The Casimir energy and its corresponding mass density will then be given by,

$$U_C(g) = \frac{G^2}{\hbar^2} m_{SUSY}^3$$

$$\rho_C(g) = \frac{U_C(g)}{\frac{4}{3}\pi\left(\frac{\lambdabar_c}{2}\right)^3 c^2} = \frac{6}{\pi}\left(\frac{G^2 c}{\hbar^5}\right) m_{SUSY}^8 \tag{11}$$

If we set $\rho_c(g)$ to the measured dark energy value of $\rho_\Lambda = 0.72 \times 10^{-29}$ g/cm$^3$, we find that

$$m_{SUSY} = \left[\frac{\pi}{6}\left(\frac{\hbar^5}{G^2 c}\right)\rho_\Lambda\right]^{1/8} \approx 5 \text{ TeV/c}^2 \tag{12}$$

Thus a SUSY-breaking energy of 5 TeV can produce a Casimir gravitational self-energy that is equal to the measured dark energy. Note that since this SUSY-breaking energy is for a virtual pair, the energy per virtual particle is 2.5 TeV.

Using the cosmological equation of energy-momentum conservation,

$$\dot{\rho} = -3(\rho + p)\frac{\dot{a}}{a} \qquad (13)$$

where $a$ is the cosmological scale factor, one can readily see that, since $\rho_c(g)$ does not vary with time in the broken-supersymmetry regime, the parameter $w$ in the equation of state, $p = w\rho$, must be = -1. The Casimir gravitational self-energy density can thus account for the measured dark energy density when the SUSY-breaking energy is $\approx$ 5 TeV, in good agreement with current estimates of the SUSY-breaking energy.

We also note that the Casimir energy density $\rho_c(g)$ can be written as,

$$\rho_C(g) = \frac{6}{\pi}\left(\frac{G^2 c}{\hbar^5}\right) m_{SUSY}^8 = \frac{6}{\pi}\left(\frac{c}{\hbar}\right)^3 \frac{m_{SUSY}^8}{m_{Pl}^4} \qquad (14)$$

Eqn. (14) is equivalent to the well-known geometric relation between the SUSY, Planck and cosmological constant, $\Lambda$, energy scales,

$$\Lambda^{1/4} \sim \frac{m_{SUSY}^8}{m_{Pl}^4} \qquad (15)$$

which, in turn, has been shown [20] to be directly related to Weinberg's anthropic principle [21], and possibly also to superstring M-theory [22]. Thus the Casimir gravitational self-energy of the vacuum fluctuations appears to provide a quantum mechanism for these important but more general relationships.

The question still remains why this Casimir gravitational self-energy of the vacuum fluctuations gravitates on a cosmological scale but the actual vacuum energy does not. Perhaps a vacuum fluctuation can gravitate only on the quantum scale because it exists in a higher (d+4) dimensional brane or because it is spread over two or more shadow branes. On the other hand, the Casimir self-energy is an energy that is imparted to the cavity, that is, to spacetime itself, resulting in an energy density of spacetime that is fully confined to our 4-dimensional brane, and thus can gravitate on a cosmological scale.

## **Acknowledgements**

The author would like to thank L. Kofman and J.R. Bond for their helpful comments and suggestions.